\documentclass[reprint, aps, prx, superscriptaddress]{revtex4-2}

\usepackage{amsmath, amssymb, graphicx}
\usepackage{xcolor}
\usepackage{physics, cancel, bm, siunitx}
\usepackage[colorlinks=true,citecolor=blue,urlcolor=blue,linkcolor=blue]{hyperref}
\usepackage{orcidlink}

\newcommand{\av}[1]{\left\langle #1 \right\rangle}

\newcommand{\meas}{\text{m}}

\newcommand{\trelax}{\tau_\mathrm{rel}}

\newcommand{\dg}{{\, \dagger}}
\newcommand{\calI}{\mathcal{I}}
\newcommand{\st}{\mathrm{eq}}

\DeclareMathOperator{\sgn}{sgn}

\begin{document}

\title{When trajectory-based bounds fail: information thermodynamics under noisy feedback}

\author{Natalia Ruiz-Pino\,\orcidlink{0000-0003-3874-4366}}
\email{nruiz1@us.es}
\affiliation{%
 Física Teórica, \href{https://institucional.us.es/mufens/}{Multidisciplinary Unit for Energy Science}, \href{https://www.us.es}{Universidad de Sevilla}, Apartado de Correos 1065, E-41080 Sevilla, Spain
}%
\affiliation{\href{https://ror.org/02feahw73}{CNRS}, \href{https://ror.org/04zmssz18}{ENS de Lyon}, \href{https://ror.org/00w5ay796}{Laboratoire de Physique}, F-69342 Lyon, France}%
\author{Ludovic Bellon\,\orcidlink{0000-0002-2499-8106}}
\email{Ludovic.Bellon@ens-lyon.fr}
\affiliation{\href{https://ror.org/02feahw73}{CNRS}, \href{https://ror.org/04zmssz18}{ENS de Lyon}, \href{https://ror.org/00w5ay796}{Laboratoire de Physique}, F-69342 Lyon, France}
\author{Antonio Prados\,\orcidlink{0000-0002-9559-8249}}
\email{prados@us.es}
\affiliation{%
 Física Teórica, \href{https://institucional.us.es/mufens/}{Multidisciplinary Unit for Energy Science}, \href{https://www.us.es}{Universidad de Sevilla}, Apartado de Correos 1065, E-41080 Sevilla, Spain
}%

\date{\today}

\begin{abstract}
Information engines exploit feedback to extract work from thermal fluctuations, extending the second law of thermodynamics through information-theoretic bounds. While several such bounds have been proposed, their relative performance under realistic conditions---where measurements are noisy and feedback is temporally correlated---remains largely unclear. Here, we experimentally and theoretically investigate this problem in an underdamped feedback-controlled system with Markovian measurements but non-Markovian control sequences. We compare three representative bounds derived from transfer entropy, unavailable information, and Markovian mutual information, and find that none is universally optimal. Instead, measurement noise preferentially affects information measures that rely on detailed trajectory statistics, while leaving quantities based on instantaneous correlations comparatively robust. As a consequence, trajectory-dependent bounds deteriorate rapidly, giving rise to a crossover in which the Markovian mutual-information bound becomes tighter than the unavailable-information bound over a broad range of measurement noise. Our results reveal a general limitation of information-theoretic descriptions that rely on detailed trajectory statistics in realistic settings and provide a unified perspective on information thermodynamics beyond idealised feedback protocols.
\end{abstract}


\maketitle


\section{Introduction}

Information can be used to extract work from thermal fluctuations, extending the second law of thermodynamics through information-theoretic bounds. A central question, however, remains: which information-theoretic description remains most thermodynamically relevant under realistic conditions? This problem lies at the heart of the thermodynamic role of information, a major theme in modern non-equilibrium physics. Since Maxwell's demon~\cite{maxwell_theory_1871} and Szilard's engine~\cite{szilard_uber_1929}, understanding the physical meaning and thermodynamic value of information has remained a central challenge in statistical physics. The seminal works of Landauer and Bennett~\cite{landauer_irreversibility_1961, bennett_thermodynamics_1982} established the modern foundations of the thermodynamics of information processing, transforming Maxwell's and Szilard's conceptual insights into a quantitative framework. This framework led to information-theoretic generalisations of the second law~\cite{sagawa_thermodynamics_2012,seifert_stochastic_2012,horowitz_second-law-like_2014,parrondo_thermodynamics_2015} and motivated a broad range of experimental studies exploring Landauer erasure and feedback-controlled information engines beyond idealised limits~\cite{toyabe_experimental_2010,berut_experimental_2012,koski_experimental_2014,paneru_lossless_2018,admon_experimental_2018,paneru_efficiency_2020,dago_dynamics_2022,archambault_inertial_2024,dago_fooling_2026,Ciampini-2025}; see also Ref.~\onlinecite{ciliberto_experiments_2017} for a review.

In feedback-controlled systems, this question has been formalised through several information-theoretic bounds relating extractable work to different descriptions of information, including transfer entropy~\cite{cao_thermodynamics_2009, horowitz_nonequilibrium_2010, sagawa_nonequilibrium_2012, abreu_thermodynamics_2012, lahiri_fluctuation_2012, admon_experimental_2018, ruiz-pino_information_2023}, unavailable information~\cite{ashida_general_2014, potts_detailed_2018, paneru_lossless_2018, archambault_inertial_2024}, and Markovian mutual information~\cite{sagawa_fluctuation_2012, horowitz_thermodynamics_2014, ruiz-pino_entropic_2026}. These bounds can be written schematically as 
\begin{equation} 
\beta \overline{W} \ge -\overline{\mathcal{I}}, 
\end{equation} 
where $\beta=(k_B T)^{-1}$, with $k_B$ the Boltzmann constant and $T$ the temperature of the thermal bath,  $\overline{W}$ denotes the average extracted work per cycle ($\overline W<0$ for work extraction), and different choices of the information term $\overline{\mathcal{I}}$ correspond to different descriptions of the information acquired and exploited by the feedback controller. Throughout this paper, all per-cycle variables are denoted with an overline. We focus on three representative information quantities: the transfer-entropy rate $\overline{\mathcal I_c}$, the unavailable-information correction $\overline{\mathcal I}-\overline{\mathcal I_u}$, and the Markovian mutual-information term $\overline{\Delta_{\meas}I}$. These quantities characterise different aspects of information processing: mutual information {$\overline{\Delta_{\meas}I}$} quantifies correlations between the system state and measurement outcomes, transfer entropy {$\overline{\mathcal I_c}$} captures directional information flow across time, and the unavailable information bound {$\overline{\mathcal I}-\overline{\mathcal I_u}
$ characterises irreversibility induced by feedback. In the limit of perfect measurements in overdamped systems, and when the measurement outcomes coincide with the system state, the unavailable-information bound is the tightest, becoming an equality for the extracted work~\cite{ashida_general_2014,paneru_lossless_2018,archambault_inertial_2024,ruiz-pino_entropic_2026}. Beyond this idealised regime, however, it remains unclear which of these information-theoretic descriptions remains the most thermodynamically relevant.

Realistic information engines generally operate outside these idealised conditions. In many experimental implementations, inertia cannot be neglected~\cite{dago_dynamics_2022, archambault_inertial_2024, archambault_information_2025, Ciampini-2025}, measurements are noisy~\cite{paneru_efficiency_2020, saha_bayesian_2022}, and feedback operations are performed at finite sampling intervals~\cite{admon_experimental_2018, paneru_efficiency_2020, archambault_inertial_2024, archambault_information_2025, dago_fooling_2026}. Even when feedback decisions depend only on the most recent measurement, corresponding to so-called Markovian measurements, the finite sampling interval induces temporal correlations that render the resulting control sequence non-Markovian. 
\begin{figure}
    \centering
    \includegraphics[width=3.25in]{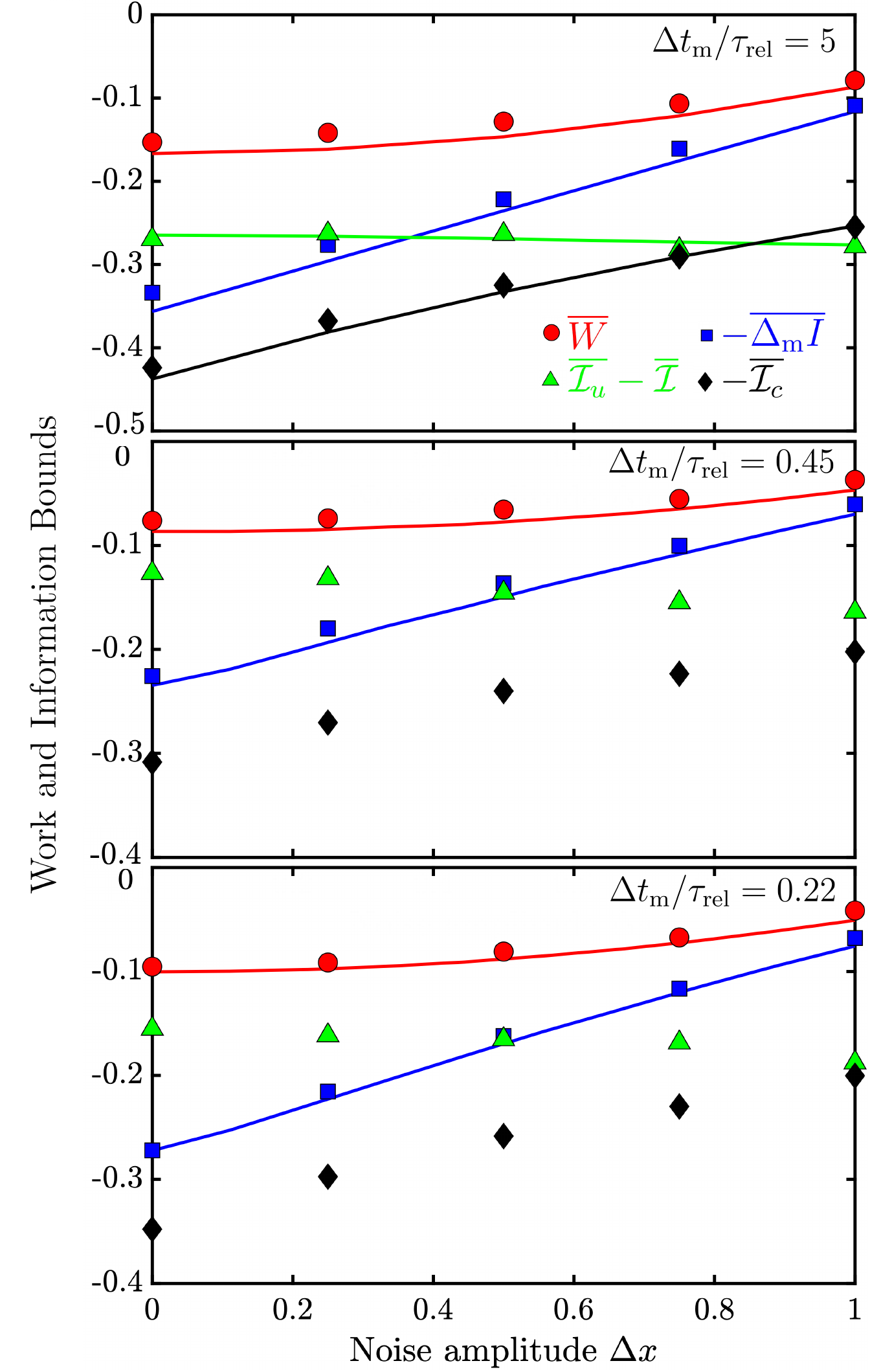}
    \caption{Comparison between the extracted work and the different information-theoretic bounds as a function of the measurement noise $\Delta x$ for $L=1$ and sampling intervals $\Delta t_{\meas}/\trelax\simeq 5$, $0.45$, and $0.22$. Extracted work in units of $k_B T$ $\beta\overline{W}$ (red), unavailable-information bound $\overline{\calI_u}[\vec{c}] - \overline{\calI}[\vec{c}]$ (green), Markovian mutual-information bound $-{\overline{\Delta_{\meas}I}[\Gamma;c]}$ (blue), and transfer-entropy bound $-\overline{\calI_c}[\vec{\Gamma}, \vec{c}]$ (black). Solid lines correspond to theoretical predictions and symbols to experimental data. The crossover between bounds as the measurement noise $\Delta x$ increases is clearly observed, revealing that trajectory-dependent and instantaneous descriptions of information exhibit markedly different sensitivities to measurement noise. Error bars, obtained by bootstrap resampling, are not visible on the scale shown.}
    \label{fig:thermodinamic_deltatm_var}
\end{figure}

Understanding the role of inertia, noise, and temporal correlations requires going beyond the standard framework of overdamped dynamics with perfect measurements and Markovian feedback. Measurement noise reduces the information available to the controller by introducing stochastic uncertainty in the measurement outcomes, whereas finite sampling intervals generate non-trivial temporal correlations in the control sequence~\cite{sagawa_nonequilibrium_2012, admon_experimental_2018, archambault_inertial_2024, paneru_efficiency_2020, ruiz-pino_information_2023, ruiz-pino_entropic_2026}. These correlations distribute information across trajectories, while measurement noise progressively limits access to the temporal structure encoded in those trajectories. As a result, different information-theoretic descriptions can exhibit markedly different sensitivities to noise and temporal correlations.

We investigate these effects in an underdamped information engine with discrete-time feedback. The system consists of an inertial particle{, characterized by its position $x$ and velocity $v$,} evolving in a controllable harmonic potential whose centre is set to $cL$, with $c=\pm1$, according to successive position measurements. By varying both the sampling interval and the measurement noise, we explore regimes ranging from near-equilibrium dynamics at long sampling intervals to strongly correlated nonequilibrium dynamics at short sampling intervals. This provides a controlled setting in which the interplay between temporal correlations and measurement noise can be systematically investigated.

Unlike many analytically tractable information engines, the system does not fully relax between consecutive measurements. Instead, it evolves towards a periodic nonequilibrium state in which the sampling interval may be much shorter than the intrinsic relaxation time $\trelax$. This regime is particularly relevant for finite-power operation and lies well outside the idealised conditions under which information-theoretic bounds are typically analysed.

A distinctive feature of the present work is that {all three information quantities} are directly evaluated in an experimentally relevant setting, despite the simultaneous presence of measurement noise and temporal correlations. This enables a quantitative comparison of the three bounds under realistic operating conditions and provides direct experimental access to information measures that are typically difficult to estimate beyond idealised regimes.

Information-theoretic quantities such as transfer entropy {$\overline{\mathcal I_c}$} and unavailable information  {$\overline{\mathcal I_u}$} depend on the full sequence of measurement outcomes and therefore rely on temporal correlations extending over multiple measurement steps. The Markovian mutual information {$\overline{\Delta_{\meas}I}$}, by contrast, is defined solely in terms of instantaneous correlations between the system state and the control variable. Comparing these bounds therefore provides a direct way of assessing how measurement noise affects trajectory-dependent and instantaneous descriptions of information differently.

Here, we employ a \textit{minimal model} in which the measurement variable $y$ is identified with the control decision: $y=c$. From an experimental perspective, this constitutes the simplest and most natural choice. The system dynamics is determined by the state $\Gamma$ and the applied protocol $c$; it does not depend on the specific choice of measurement variable. The evaluation of transfer entropy and unavailable information, however, requires probability distributions defined over long sequences of states $\vec{\Gamma}_k\equiv\{\Gamma_1, \ldots, \Gamma_k\}$ and measurement outcomes $\vec{y}_k\equiv\{y_1, \ldots, y_k\}$. Any alternative choice for $\vec{y}_k$ would therefore substantially complicate the estimation of these quantities without altering the underlying dynamics.

We show that no single bound is universally optimal. Instead, their relative performance is governed by how measurement noise affects the underlying information-theoretic description. Our central result, summarised in Fig.~\ref{fig:thermodinamic_deltatm_var}, is that measurement noise differentially affects trajectory-dependent and instantaneous descriptions of information, leading to qualitatively different thermodynamic relevance under realistic operating conditions. 

In the limit of perfect measurements, unavailable information {$\overline{\mathcal I_u}-\overline{\mathcal I}$} provides the tightest bound, although it does not coincide with the extracted work. As the measurement noise increases, bounds that rely on detailed trajectory statistics deteriorate rapidly, with the unavailable-information bound exhibiting the strongest degradation. Therefore, the simpler Markovian mutual-information bound {-$\overline{\Delta_{\meas}I}$} becomes the tightest over a broad range of measurement noise, even though its definition does not incorporate temporal correlations. Moreover, its sign change closely tracks the disappearance of work extraction---see Fig. S2 in Supplementary Information, correctly identifying the regime in which the system ceases to operate as an information engine. The transfer-entropy bound {-$\overline{\mathcal I_c}$} remains looser than the Markovian mutual-information one, consistent with general properties previously established in overdamped systems~\cite{ruiz-pino_entropic_2026} and extended here to the underdamped regime in the Supplementary Information. Remarkably, at sufficiently large noise, the transfer-entropy bound can still outperform the unavailable-information bound, highlighting the distinct ways in which measurement noise affects different trajectory-dependent descriptions.

Overall, our results show that the thermodynamic value of information is inherently context dependent and cannot be fully captured by a single information measure. More broadly, the present study provides an operational perspective on information thermodynamics beyond memoryless feedback control, revealing how measurement noise limits the thermodynamic relevance of information-theoretic descriptions that rely on increasingly detailed trajectory statistics, while leaving instantaneous correlations comparatively robust. This perspective is directly relevant to a wide range of experimental systems in which inertia, noise, and temporal correlations are intrinsic and therefore unavoidable.


\section{Information-theoretic bounds for the extractable work}

Information engines operate through a feedback controller (or demon) that measures the system state and subsequently applies a protocol conditioned on the measurement outcome. When the control decision depends only on the most recent measurement outcome, the feedback decision is said to be Markovian. Throughout this work, measurements are performed periodically at times $t_n=n\Delta t_{\meas}$, with Markovian feedback decisions. The feedback loop is naturally described by three variables: the system state $\Gamma$, the measurement outcome $y$ generated from $\Gamma$, and the applied protocol $c=c(y)$, taken here as a deterministic function of the measurement outcome.
Here, $n$ denotes the measurement index, while $k$ denotes the total number of measurements.

\begin{figure}
    \centering
    \includegraphics[width=3.375in]{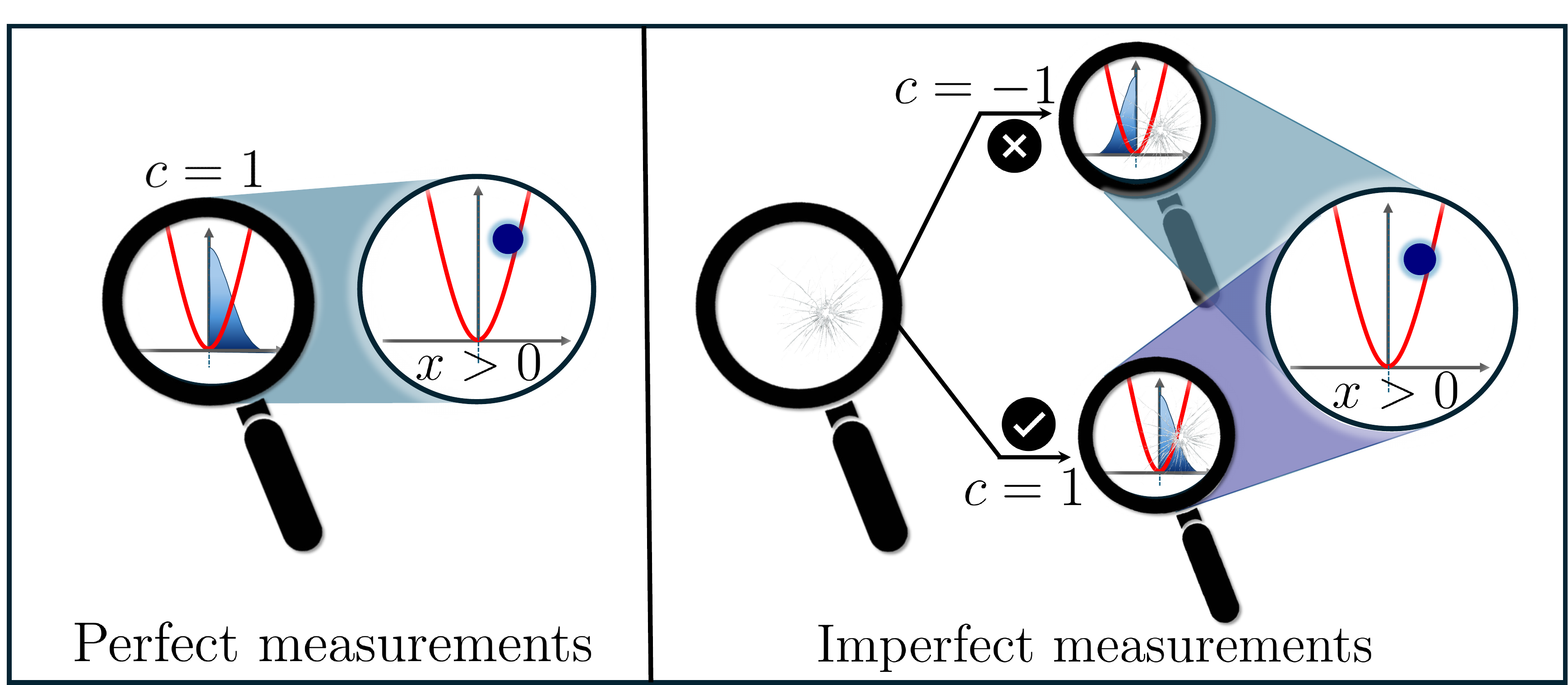}
    \caption{Schematic representation of perfect and imperfect measurements for a Brownian particle in a harmonic trap. The particle position $x$ is probed by a binary detector that identifies whether it lies to the right ($c=1$) or left ($c=-1$) of the trap centre. For perfect measurements (left), the outcome is uniquely determined by the particle position. For imperfect measurements (right), measurement uncertainty allows the same position $x$ to yield either outcome, consistent with the conditional probability $\Theta(c|x)$ defining the measurement model.}
    \label{fig:perfect-imperfect}
\end{figure}
Measurements may be either perfect or imperfect, as illustrated in Fig.~\ref{fig:perfect-imperfect}. Perfect measurements partition phase space into non-overlapping regions $\{\chi_y\}$ such that, whenever $\Gamma\in\chi_y$, the outcome $y$ is obtained with unit probability. They are therefore characterised by the conditional probability $\Theta(y|\Gamma)=\mathbf{1}_{\chi_y}(\Gamma)$, where $\mathbf{1}_{\chi_y}(\Gamma)$ denotes the indicator function of the region $\chi_y$. Perfect measurements thus define a coarse-graining of phase space, with different choices of $\chi_y$ corresponding to different measurement resolutions. Imperfect measurements are affected by intrinsic uncertainty that introduces stochastic variability in the measurement outcomes $y$. In this case, $\Theta(y|\Gamma)\neq \mathbf{1}_{\chi_y}(\Gamma)$, with $\Theta(y|\Gamma)$ taking values in the interval $[0, 1]$.

In this framework, energetic quantities such as work and heat depend only on the system state $\Gamma$ and the applied protocol $c=c(y)$. The entropic balance instead depends explicitly on the choice of measurement variable $y$, since entropy is inherently description dependent. Consequently, different information measures lead to different refinements of the second law.

For feedback-controlled systems, Sagawa and Ueda~\cite{sagawa_nonequilibrium_2012} generalised the second law by introducing the transfer entropy
\begin{equation}\label{eq:Ic-Gamma-y-def}
\mathcal I_c[\vec{\Gamma}_k, \vec y_k]
=
\left\langle
\ln
\frac{\Theta(\vec y_k|\vec\Gamma_k)}
{P(\vec y_k)}
\right\rangle, 
\end{equation}
where $\vec{\Gamma}_k\equiv\{\Gamma_1, \ldots, \Gamma_k\}$ and $
\vec y_k\equiv\{y_1, \ldots, y_k\}$
denote the chains of system states and measurement outcomes at the measurement times, up to a certain time $t_k=k\Delta t_{\meas}$, $\Theta(\vec{y}_k|\vec{\Gamma}_k)\equiv \prod_{n=1}^k \Theta(y_n|\Gamma_n)$, and the average is taken over all possible trajectories. The corresponding refinement of the second law reads
\begin{equation}\label{eq:2nd-law-sagawa}
    \beta(\av{W}-\Delta F)
    \ge
    -\mathcal I_c[\vec{\Gamma}_k, \vec{y}_k], 
\end{equation}
where  $\av{W}$ denotes the average work, and $\Delta F$ the free-energy difference. 

\begin{figure*}
    \centering
    \includegraphics[width=6in]{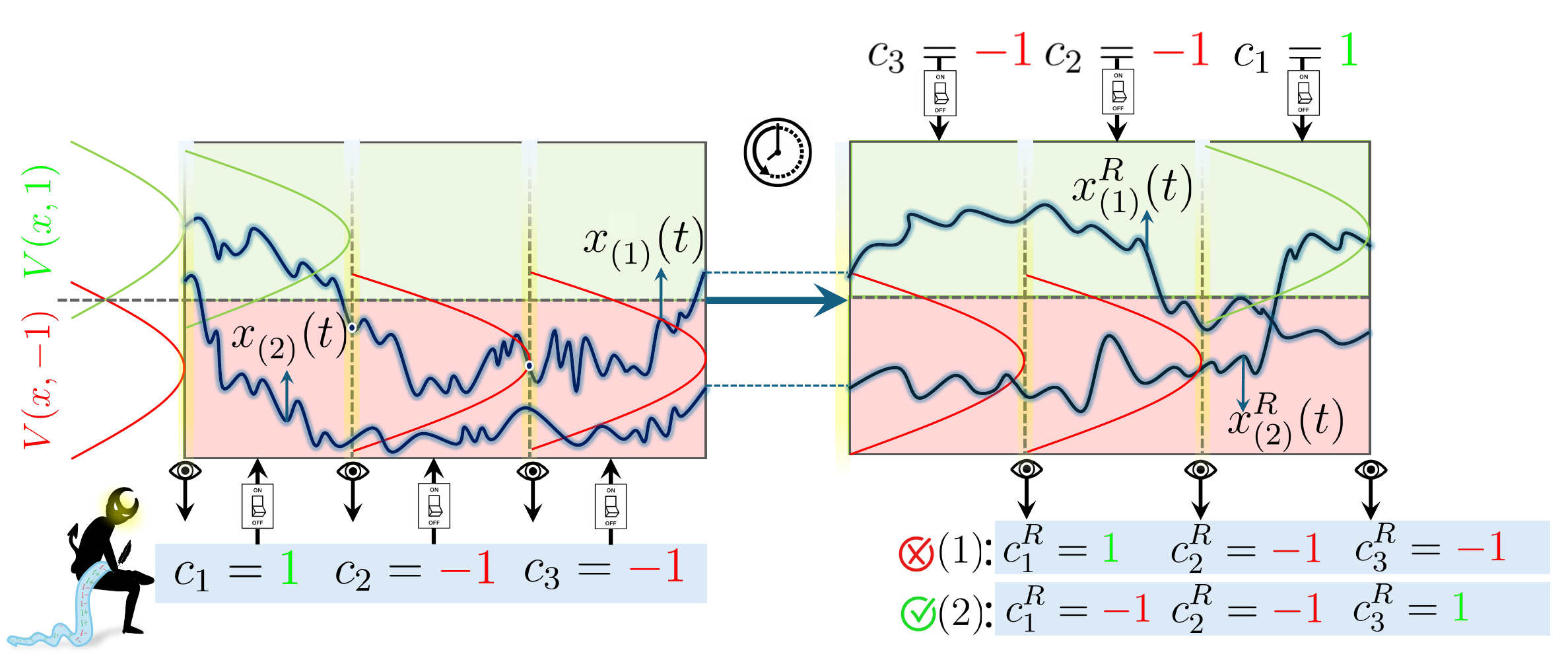}
    \caption{Forward and backward processes for a feedback-controlled information engine, illustrated for an overdamped particle with $\Gamma=\{x\}$ and measurement outcome $y=c$. Left: in the forward process, the controller measures the particle position at discrete times and updates the potential $V(x, c)$ accordingly, generating a measurement sequence $\vec y$. Right: in the backward process, the time-reversed sequence $\vec y^\dg$ is applied without feedback (open-loop protocol), while measurements are still performed, yielding a backward sequence $\vec y^R$. The unavailable information is determined by the probability that the backward sequence reproduces the time-reversed forward one, $\vec y^R=\vec y^\dg$. In the example shown, only the trajectory $x^R_{(2)}$ contributes to the unavailable information.}
    \label{fig:explanation}
\end{figure*}
Although Eq.~\eqref{eq:2nd-law-sagawa} applies to both perfect and imperfect measurements, the transfer entropy diverges when the measurement variable is identified with the system state, $y=\Gamma$. To overcome this limitation, Ashida \textit{et al.}~\cite{ashida_general_2014} introduced the unavailable information for perfect measurements, which was later extended to imperfect measurements by Potts~\cite{potts_detailed_2018}. This quantity is defined in terms of the backward process associated with the control sequence generated during the forward dynamics. The construction of the corresponding backward process is illustrated in Fig.~\ref{fig:explanation} for the particular case $c=y$. 

Denoting by $\vec y^\dg$ the time-reversed measurement sequence and by $\vec y^R$ the measurement outcomes obtained during the backward evolution, the unavailable-information bound becomes
\begin{subequations}\label{eq:2nd-law-Ash}
   \begin{align}
   \beta(\langle W\rangle-\Delta F)
    &\ge
    \calI_u[\vec{y}_k]-\calI[\vec{y}_k], \\
    \calI[\vec{y}_k]
    &\equiv
    -\av{\ln P(\vec{y}_k)}, \\ 
    \calI_u[\vec{y}_k]
    &\equiv
    -\av{
    \ln
    P^R(\vec{y}_k^R=\vec{y}_k^\dg|\vec{y}_k^\dg)
    }.
\end{align} 
\end{subequations}
Here, $\calI[\vec{y}_k]$ is the entropy of the chain of measurements and $\calI_u[\vec{y}_k]$ is the so-called unavailable information~\cite{ashida_general_2014, potts_detailed_2018}. Unlike transfer entropy, unavailable information remains well defined when the measurement outcome is identified with the system state, $y=\Gamma$. In this limit, it possesses a particularly remarkable property: inequality~\eqref{eq:2nd-law-Ash} becomes saturated in overdamped systems~\cite{ruiz-pino_entropic_2026}. In the Supplementary Information, we extend this result to the underdamped regime. Evaluating the unavailable information requires detailed knowledge of the backward process, which generally makes its experimental and numerical estimation challenging.

For the class of information engines considered here, the joint stochastic process $(\Gamma, y)$ is Markovian for all times $t$, where $\Gamma=\{x, v\}$ in the underdamped regime studied throughout this work. Furthermore, the system reaches a long-time periodic state with period $\Delta t_{\meas}$, and {the distributions of} both $\Gamma_n$ and $y_n$ become stationary over the coarse-grained timescale defined by the measurement times $t_n=n\Delta t_{\meas}$. Remarkably, although the chain of system states remains Markovian, the corresponding chain of measurement outcomes is generally non-Markovian. 

A third refinement of the second law then follows from the Markovian character of the joint process $(\Gamma, y)$, 
\begin{subequations}\label{eq:2nd-law-gener-deltamI} 
    \begin{align}
        \beta \left(\av{W}-\Delta F\right)&\geq -{\Delta_{\meas}I[\vec{\Gamma}_k;\vec{y}_k]}, \\
        \Delta_{\meas} I[\vec{\Gamma}_k;\vec{y}_k]
        &=I^+[\vec{\Gamma}_k;\vec{y}_k]-I^-[\vec{\Gamma}_k;\vec{y}_k].
    \end{align} 
\end{subequations}
The pre- and post-measurement mutual information up to the $k$-th measurement are given by
\begin{equation}\label{eq:def-I-mutua}
I^\pm[\vec{\Gamma}_k;\vec{y}_k]
=
\sum_{n=1}^k
\left\langle
\ln
\frac{P(y|\Gamma, t_n^\pm)}{P(y, t_n^\pm)}
\right\rangle .
\end{equation}
Here, the post-measurement conditional probability satisfies $P(y|\Gamma, t_n^+)\equiv \Theta(y|\Gamma)$, as defined by the measurement model, while $P(y|\Gamma, t_n^-)$ denotes the corresponding pre-measurement conditional probability. The overdamped version of these results was established in Refs.~\cite{ruiz-pino_markovian_2024, ruiz-pino_entropic_2026}; the extension to underdamped dynamics is presented in the Supplementary Information.

In the long-time periodic state, the extracted work increases linearly with the number of measurements $k$, making the extracted work per cycle, 
$\overline W =\lim_{k\to\infty} \av{W}/k$, 
the relevant thermodynamic quantity. The different information-theoretic bounds can therefore be compared through their asymptotic rates per feedback cycle, 
$\overline f\equiv\lim_{k\to\infty}{f}/{k}$. In this limit, the trajectory variables $\vec{\Gamma}_k$ and $\vec{y}_k$ are understood as arbitrarily long sequences, and the subscript $k$ is therefore omitted in overline quantities.

From the three generalisations of the second law we have put forward, Eqs.~\eqref{eq:2nd-law-sagawa}, \eqref{eq:2nd-law-Ash}, and \eqref{eq:2nd-law-gener-deltamI}, in the limit of many measurements, the following bounds for the extractable work follow:
\begin{subequations}\label{eq:all-bounds-together}
  \begin{align}
    \beta\overline{W} \ge & 
        - {\overline{\calI_c}[\vec{\Gamma}, \vec{y}]}, \\  \beta\overline{W} \ge & \;{\overline{\calI_u}[\vec{y}] - \overline{\calI}[\vec{y}]} , \\ \beta\overline{W} \ge & 
        -  {\overline{\Delta_{\meas}I}[\Gamma;y]}, 
\end{align}  
\end{subequations}
where we have used that $\Delta F\to 0$ due to the periodicity. Our notation stresses that $\overline{\calI_c}$, $\overline{\calI_u}$  and $\overline{\calI}$ depend on the full measurement and control sequences, whereas $\overline{\Delta_{\meas}I}$ only depends on the instantaneous pre- and post-measurement states: Equation~\eqref{eq:def-I-mutua} simplifies to
\begin{equation}\label{eq:def-I-mutua-v2}
I^\pm[\Gamma;y] \equiv I^\pm[\Gamma;y, t_n^\pm]=\av{ \ln \frac{P(y|\Gamma, t_n^\pm)}{P(y, t_n)}} , 
\end{equation}
which is independent of $n$.

At the level of the second-law inequalities in Eqs.~\eqref{eq:all-bounds-together}, the three bounds remain formally valid for arbitrary measurements. The only limitation is that transfer entropy and Markovian mutual information become ill-defined when the measurement outcome is identified with the full system state, $y=\Gamma$, owing to delta-like divergences in the corresponding conditional probabilities. In the minimal model considered throughout this work, however, the measurement outcome coincides with the control parameter, $y=c$, such that all three bounds remain well defined for both perfect and imperfect measurements.

For perfect measurements, the transfer entropy reduces to the entropy of the measurement sequence, $\calI_c[\vec{\Gamma}_k, \vec{c}_k]=\calI[\vec{c}_k]$, 
which leads to the ordering
\begin{equation}\label{eq:transfer-unav-ordering}
    \beta\overline{W}
    \ge
    \overline{\calI_u}[\vec{c}]-\overline{\calI}[\vec{c}]
    \ge
    -\overline{\calI_c}[\vec{\Gamma}, \vec{c}].
\end{equation}
By contrast, this ordering generally breaks down in the presence of imperfect measurements, where measurement noise generates additional contributions to the transfer entropy---see Methods.

A more robust relation exists between the transfer-entropy and Markovian mutual-information bounds. For both perfect and imperfect measurements, 
\begin{equation}
    \beta\overline{W}
    \ge
    -\overline{\Delta_{\meas}I}[{\Gamma};{c}]
    \ge
    -\overline{\calI_c}[\vec{\Gamma}, \vec{c}], 
\end{equation}
so that the Markovian bound is always tighter than the transfer-entropy one~\cite{ruiz-pino_entropic_2026}; the extension to underdamped dynamics is presented in the Supplementary Information. No general ordering, however, is known between the unavailable-information and Markovian mutual information bounds. Determining their relative performance in the presence of noisy and temporally correlated feedback constitutes one of the central goals of the present work.


\section{Underdamped information engine}\label{sec:experimental-setup}

We experimentally investigate an underdamped information engine based on a conductive micro-cantilever clamped above a plane electrode~\cite{archambault_inertial_2024, archambault_information_2025}. The motion of the cantilever is monitored by interferometry at a position dominated by the fundamental flexural mode, allowing the system to be modelled as an underdamped harmonic oscillator with state $\Gamma=(x, v)$. {The oscillator angular resonance frequency is $\omega_0 = 2\pi\times ( \SI{1100}{Hz})$ and the quality factor is $Q_f \simeq 7$ under ambient conditions at temperature $T=\SI{295}{K}$. The relaxation time of the oscillator is $\trelax=2Q_f/\omega_0=\SI{2}{ms}$. Throughout this work, lengths are expressed in units of the equilibrium positional fluctuations amplitude $\sigma=\sqrt{\av{x^2}}=\sqrt{k_B T/k}\simeq \SI{1}{nm}$ (with $k$ the stiffness of the cantilever), times in units of $\omega_0^{-1}$, and energies in units of $k_B T$.} Dimensionless units are used henceforth unless stated otherwise. 

A controllable electrostatic force generated by the applied voltage between cantilever and electrode produces a harmonic potential $V(x, c)=\frac{1}{2}(x-cL)^2$~\cite{Dago-2022-JStat,Dago-chapter}, whose centre is shifted according to the feedback variable $c=\pm1$. At discrete times $t_n=n\Delta t_{\meas}$, the cantilever position is measured and the control parameter is updated from the sign of the measured position. When $x_n\equiv x(t_n)<0$, the protocol $c_n=-1$ is applied, whereas for $x_n>0$ the protocol becomes $c_n=+1$, i.e. $c_n=\sgn(x_n)$. This feedback systematically lowers the oscillator energy, resulting in work extraction. Similar protocols have been studied in information engines where measurements are used to bias the system towards configurations that enhance work extraction \cite{toyabe_experimental_2010, abreu_thermodynamics_2012, ashida_general_2014, toyabe_nonequilibrium_2015, paneru_lossless_2018, paneru_efficiency_2020}.

{The interferometer sensitivity is better than $\SI{1}{pm}\simeq 10^{-3}\sigma$~\cite{paolino_quadrature_2013}, making the measurement of the position $x$ intrinsically very close to perfect. The feedback loop purposely induces imperfect measurements by adding to $x$ a flat noise $\eta$ of controlled amplitude $2\Delta x$, before applying the sign function to determine the value of $c$: $c_n=\sgn(x_n+\eta_n)$. With this implementation of the information engine, the demon only accesses a noisy estimate of the particle position, modelled through the conditional probability $\Theta(c_n|x_n)$: $\Theta(+1|x_n)$ and $\Theta(-1|x_n)$ are proportional to the length of the interval $[x_n-\Delta x, x_n+\Delta x]$ to the right and to the left of the origin, respectively. This approach reproduces the scheme illustrated in Fig.~\ref{fig:perfect-imperfect}, with a tunable uncertainty $\Delta x$.}

Although the dynamics is underdamped, the feedback decision depends only on position. Therefore, in our minimal model, where the measurement outcome coincides with the control variable, $y=c$, the general conditional probability $\Theta(y|\Gamma)$ reduces to $\Theta(c|x)$. Consequently, post-measurement quantities, such as $I[\Gamma;c, t_n^+]$ and $\overline{\calI_c}[\vec{\Gamma}, \vec{c}]$, depend only on $x$, whereas pre-measurement quantities such as $I[\Gamma;c, t_n^-]$ retain an explicit dependence on the velocity degree of freedom.


\section{Results}\label{sec:results}

\begin{figure}[t]
    \centering
    \includegraphics[width=3in]{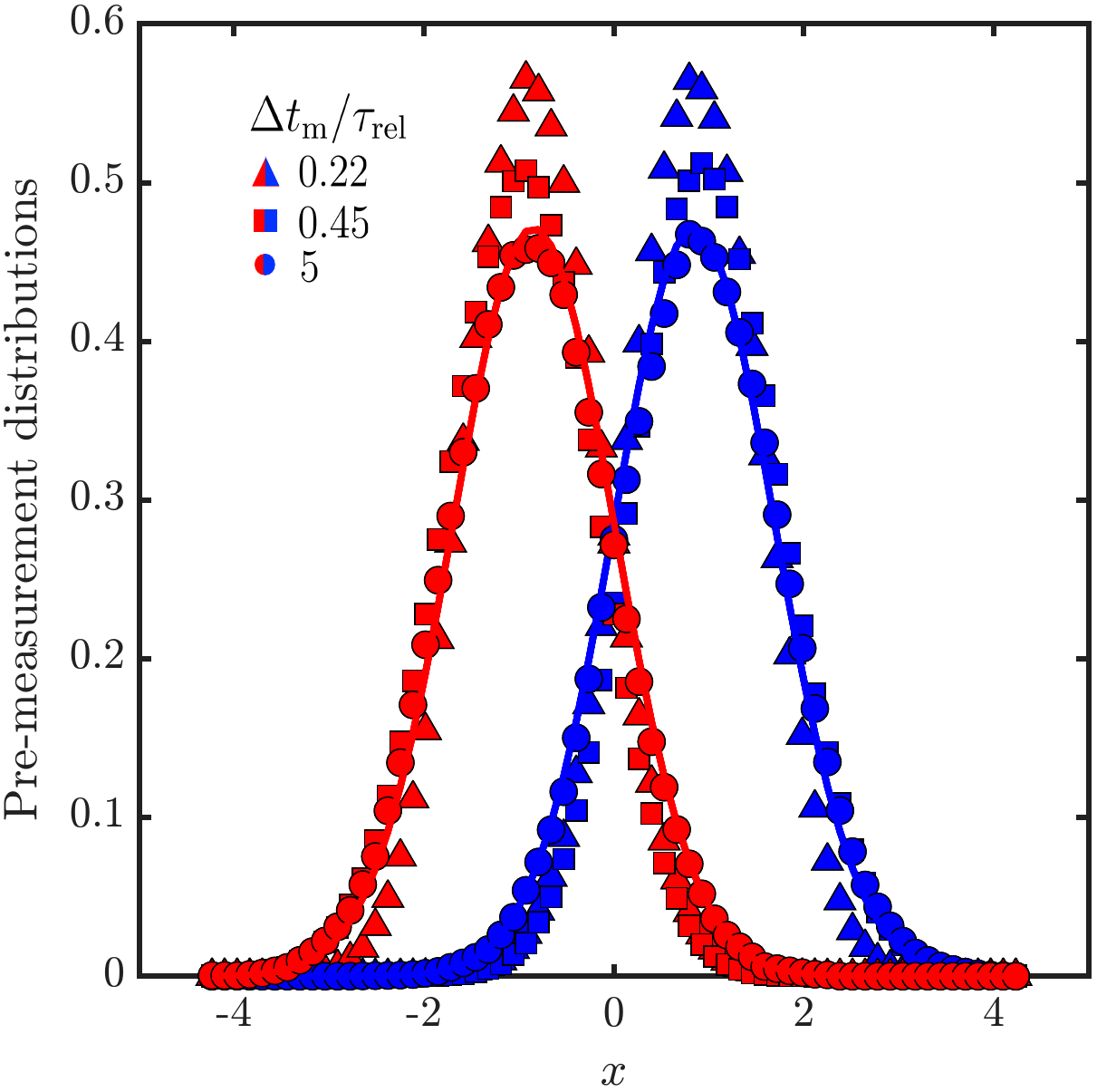}
    \caption{Position probability distributions immediately before the measurement {for $L=1$ and sampling intervals $\Delta t_{\meas}/\trelax=5$, $0.45$, and $0.22$}. Symbols show the experimental estimate of $P(x|c, t_n^-)$ in the long-time periodic state, while solid lines denote the corresponding equilibrium distribution $P_{\mathrm{eq}}(x|c)=\exp[-(x-cL)^2/2]/\sqrt{2\pi}$. Blue and red correspond to $c=+1$ and $c=-1$, respectively. For the longest $\Delta t_{\meas}$, the system relaxes between measurements and reaches equilibrium, whereas for shorter $\Delta t_{\meas}$ the distributions remain out of equilibrium. This deviation from equilibrium reflects the emergence of temporal correlations in the control sequence.}
    \label{fig:pdf_vs_peq}
\end{figure}
We now compare the extracted work with the three information-theoretic bounds: transfer entropy, unavailable information, and Markovian mutual information. By varying both the sampling interval $\Delta t_{\meas}$ and the measurement uncertainty $\Delta x$, we investigate how temporal correlations and measurement noise affect the relative performance of these bounds.

\subsection{Memory effects and estimation of information rates}\label{sec:bounds-estimators}

Figure~\ref{fig:pdf_vs_peq} characterises the different dynamical regimes accessed in the experiment through the pre-measurement probability distribution $P(x|c, t_n^-)$. For long sampling intervals, $\Delta t_{\meas}\simeq 5\trelax$, the system relaxes to the equilibrium distribution associated with the applied protocol between consecutive measurements, which corresponds to the regime most commonly considered in the literature. For shorter sampling intervals $\Delta t_{\meas} <\trelax$, the system remains out of equilibrium, generating temporal correlations in the control sequence and thus non-Markovian control dynamics.

The extracted work and the Markovian mutual-information bound depend only on the instantaneous values $(\Gamma_n, c_n)$ and can therefore be computed directly from the corresponding stationary distributions and propagators. The transfer-entropy and unavailable-information bounds, instead, depend on the full control sequence $\vec c_k$, requiring dedicated finite-memory estimators.

The main challenge for comparing the different thermodynamic bounds arises from the fact that, although the measurements are Markovian, the control sequence $\vec{c}_k$ is not. As a consequence, evaluating quantities such as the entropy rate ${\cal I}[\vec{c}_k]$ and the unavailable-information rate ${\calI_u}[\vec{c}_k]$ requires handling temporal correlations in the control sequence.

Our approach relies on the observation that these correlations remain finite-ranged{, since $\trelax$ itself is finite}. The non-Markovian control sequence can therefore be accurately described through finite-memory estimators characterised by memory length $M$. In the long-time periodic state, both information rates converge to constants, $\overline{\calI}[\vec c]\equiv \lim_{k\to\infty}{\calI[\vec{c}_k]}/{k}=\mathcal H$ and $\overline{\calI_u}[\vec c]\equiv \lim_{k\to\infty}{\calI_u[\vec{c}_k]}/{k}=\mathcal J$, which can be estimated directly from experimental trajectories. 
\begin{figure}
     \centering
     \includegraphics[width=3.375in]{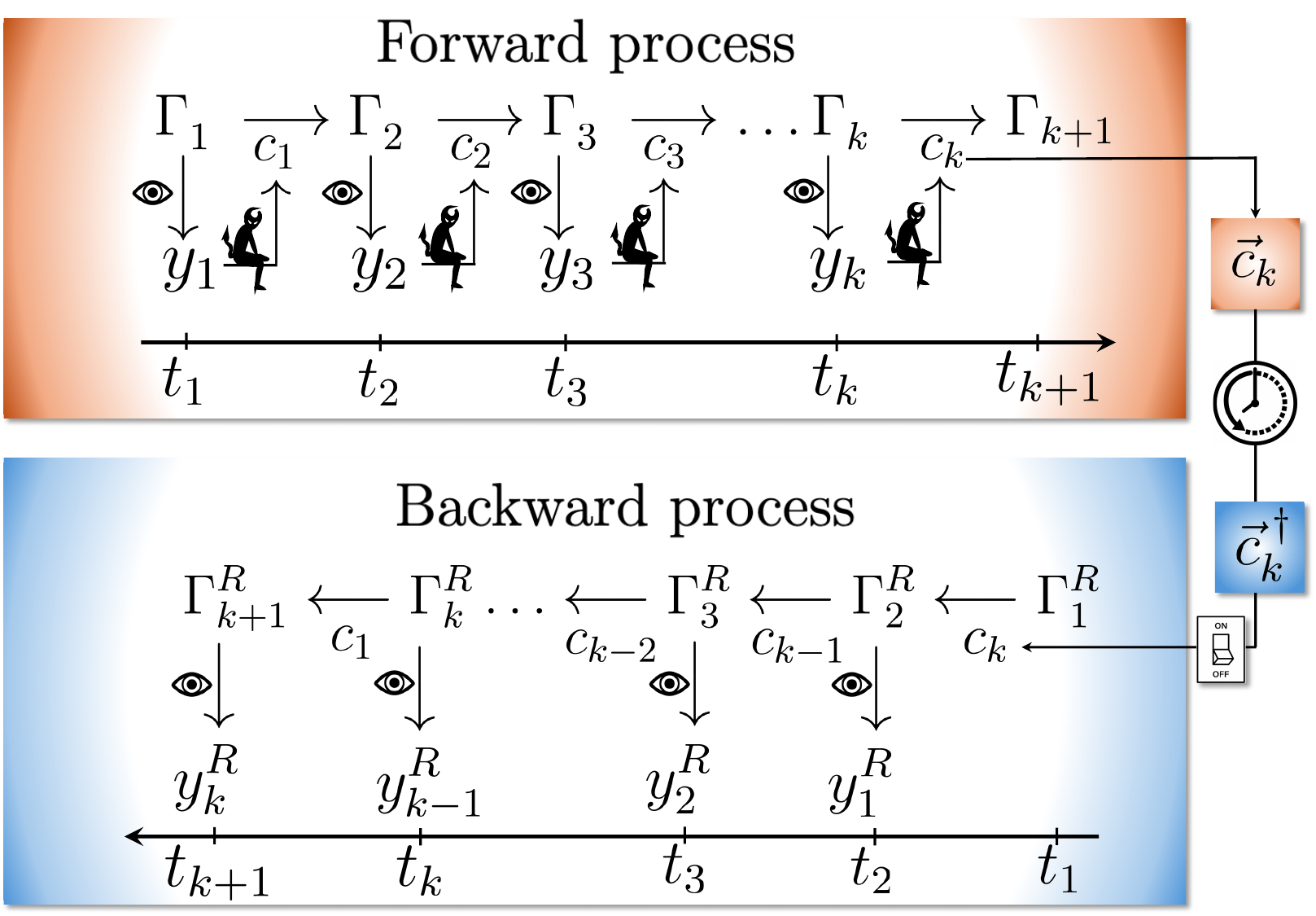}
     \caption{Schematic representation of the forward and backward processes. In the forward process, the system state $\Gamma_n$ is measured at each time $t_n$, yielding the outcome $y_n$, which determines the protocol $c_n$. The resulting control sequence $\vec c_k$ defines a backward process in which the time-reversed sequence $\vec c_k^\dg$ is applied without feedback. Measurements $y_n^R$ are still performed during the backward evolution, but do not affect the applied protocol.}
     \label{fig:back_for_processes}
\end{figure}

\begin{figure}[t]
    \centering
    \includegraphics[width=3in]{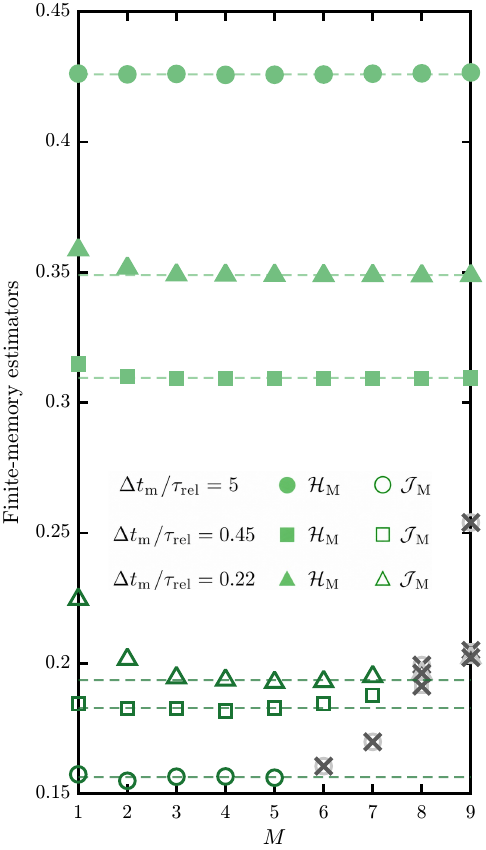}
    \caption{Convergence of the finite-memory estimators for $\overline{\calI}[\vec c]$ and $\overline{\calI_u}[\vec c]$. Entropy-rate estimator $\mathcal H_M$ (solid markers) and corresponding unavailable-information estimator $\mathcal J_M$ (open markers)  as a function of the assumed memory length $M$ for different sampling intervals $\Delta t_{\meas}$. In both cases, clear plateaux are observed, indicating that temporal correlations in the control sequence are effectively finite-ranged. Crossed-out symbols mark data points excluded due to insufficient statistics.}
    \label{fig:J_H_convergence}
\end{figure}
The unavailable information requires a backward experiment constructed from the time-reversed control sequence, illustrated in Fig.~\ref{fig:back_for_processes}. In this backward process, measurements are performed but do not affect the applied protocol. The unavailable-information rate is then determined from the probability that the backward measurement sequence reproduces the time reversal of the forward one---see also Fig.~\ref{fig:explanation}.

In practice, temporal correlations can be truncated at a finite memory length $M$. This truncation gives rise to finite-memory estimators ${\cal H}_M$ and ${\cal J}_M$, which only keep the conditioning on the previous $M$ control actions in the sequence. Increasing $M$ therefore probes progressively longer temporal correlations. For large enough $M$, one has $\mathcal{H}_M \to \mathcal{H}$ and $\mathcal{J}_M \to \mathcal{J}$.}

Figure~\ref{fig:J_H_convergence} shows the convergence of the estimators $\mathcal H_M$ (for $\overline{\calI}[\vec c]$) and $\mathcal J_M$ (for $\overline{\calI_u}[\vec c]$) as a function of the memory length $M$. Both estimators rapidly approach well-defined plateaux, demonstrating that the feedback dynamics can be accurately described within a finite-memory framework despite the non-Markovian character of the control sequence. The effective memory decreases as the sampling interval $\Delta t_{\meas}$ increases, and the control sequence becomes effectively Markovian in the long-sampling-time limit. Accurate estimation of both $\mathcal{H}_M$ and $\mathcal{J}_M$ is only possible for $M \lesssim M_{\max}$, where $M_{\max}$ depends on their respective sample sizes. For $M \gtrsim M_{\max}$ (crossed-out points), the sample becomes too small compared to the size ($2^{M+1}$) of the space of possible control sequences of length $M+1$.

The detailed derivation, numerical implementation (including sample sizes used), and statistical analysis of the estimators are presented in the Methods and Supplementary Information.


\subsection{Comparison between thermodynamic bounds}

We now compare the extracted work with the three information-theoretic bounds across both equilibrium and non-equilibrium regimes.


\subsubsection{Long sampling intervals}

We first analyse the limit $\Delta t_{\meas}\gg\trelax$, corresponding to the equilibrium regime shown in the top panel of Fig.~\ref{fig:thermodinamic_deltatm_var}. In this limit, the system relaxes to the equilibrium state associated with the applied protocol between consecutive measurements. Consequently, the forward and backward control sequences become effectively Markovian, considerably simplifying the evaluation of the chain-dependent bounds. Analytical expressions for the transfer entropy and unavailable information can then be derived directly from the corresponding transition probabilities (see Methods).

Despite the long time between measurements, consecutive control actions remain correlated, such that the control sequence is Markovian but not independent. In this regime, the information rates reduce to conditional entropies and conditional backward probabilities, allowing a direct comparison between the three second-law bounds.

For nearly perfect measurements, $\Delta x\to0^+$, it is the unavailable-information bound that provides the tightest constraint. This behaviour is consistent with the fact that unavailable information becomes optimal for perfect measurements, although in the present minimal model the choice $y=c$ prevents exact saturation of the extracted work.


\subsubsection{Finite sampling intervals}

We now turn to the experimentally relevant regime of finite sampling intervals, corresponding to the middle and bottom panels of Fig.~\ref{fig:thermodinamic_deltatm_var}. In this case, the system does not relax between consecutive measurements and the control sequence $\vec c_k$ becomes genuinely non-Markovian in both the forward and backward processes. The transfer-entropy and unavailable-information bounds therefore depend explicitly on temporal correlations extending over multiple measurement steps, requiring finite-memory estimators capable of accounting for non-Markovian feedback dynamics.

As the measurement noise increases, trajectory-based bounds progressively deteriorate, with the unavailable-information bound exhibiting the strongest degradation. This behaviour originates from the faster growth of $\overline{\mathcal I}[\vec c]$ relative to $\overline{\mathcal I_u}[\vec c]$, as shown in the Supplementary Information. Consequently, the unavailable-information correction becomes progressively less effective as the measurements become noisier.

Both the transfer-entropy and Markovian mutual-information bounds explicitly incorporate the measurement uncertainty through the conditional probability $\Theta(c|x)$, therefore capturing more accurately the degradation of the extracted work at finite noise. In particular, the Markovian mutual-information bound systematically outperforms the transfer-entropy one, in agreement with the general ordering relations discussed above.

Most notably, the Markovian bound becomes the tightest of the three over a broad range of measurement uncertainties, even though its definition does not incorporate temporal correlations in the control sequence. This shows that incorporating detailed non-Markovian information into a thermodynamic bound does not necessarily improve its tightness in noisy regimes, because the trajectory statistics on which such descriptions rely are more strongly affected by measurement noise than the corresponding instantaneous correlations.

The comparison between different sampling intervals further reveals the role of temporal correlations. As $\Delta t_{\meas}$ decreases, the control sequence becomes increasingly non-Markovian, enhancing the discrepancy between the different bounds. Nevertheless, the qualitative crossover between unavailable-information dominance at low noise and Markovian-mutual-information dominance at high noise remains robust across all investigated regimes.

Overall, these results demonstrate that the relative performance of information-theoretic bounds is governed by how measurement noise affects different information-theoretic descriptions. In particular, descriptions that rely on detailed trajectory statistics deteriorate much faster than those based on instantaneous correlations, leading to the observed crossover between bounds.


\section{Discussion}

Our results provide a unified picture of information thermodynamics under realistic operating conditions, where finite sampling intervals simultaneously give rise to finite-power operation and temporal correlations, while measurement noise limits the effective use of the resulting information. In particular, they show that the thermodynamic relevance of different information-theoretic descriptions is not universal, but depends on how they respond to measurement noise in the presence of those correlations.

This behaviour reflects a clear physical mechanism. Measurement noise affects quantities that rely on detailed trajectory statistics more strongly than those based on instantaneous correlations, thereby reducing the effectiveness of bounds that require increasingly resolved temporal information. As a result, simpler bounds based on instantaneous quantities can become tighter in noisy regimes, even though they do not explicitly account for temporal correlations.

These results demonstrate that increasingly detailed trajectory-dependent descriptions do not necessarily improve thermodynamic bounds. When the trajectory statistics on which those descriptions rely become progressively degraded by measurement noise, simpler descriptions based on instantaneous correlations can become thermodynamically more relevant.

An important aspect of the present framework is the distinction between coarse-graining and measurement noise. In this work, the coarse-graining is fixed by the choice of measurement variable and determines the level of description of the system, while measurement noise introduces an additional stochastic degradation of information within this description. Although different choices of coarse-graining could quantitatively affect the values of the information measures, the effects identified here arise from the way measurement noise degrades access to the temporal structure encoded in correlated trajectories within a fixed level of description. The observed crossover therefore does not originate from changing the coarse-graining itself.

Temporal correlations and measurement noise play distinct roles. Temporal correlations distribute information across extended measurement sequences, making quantities such as transfer entropy and unavailable information sensitive to trajectory-dependent statistics. Measurement noise, by contrast, selectively limits access to the temporal structure encoded in those sequences.

Our results also clarify the role of unavailable information as a thermodynamic bound. Originally introduced to account for absolute irreversibility and singular measurements~\cite{ashida_general_2014}, unavailable information has proven exceptionally successful in idealised scenarios involving perfect measurements~\footnote{Singular measurements assign zero probability to certain regions of phase space. Namely, there exists at least one value of $y$ such that $\Theta(y|\Gamma)=0$ for $\Gamma\notin\chi_y$. Non-singular measurements do not generate regions of phase space with strictly zero probability for a given outcome $y$.}. The present results show that this favourable behaviour does not generally survive under realistic noisy conditions. Although unavailable information remains the tightest bound in the appropriate perfect-measurement limit, it deteriorates rapidly as measurement uncertainty increases, eventually becoming less thermodynamically relevant than considerably simpler quantities. This highlights the distinction between a quantity that is optimal in idealised settings and one that remains operationally useful in practical implementations.

An additional contribution of the present work is methodological. Quantities such as transfer entropy and unavailable information provide some of the most detailed descriptions of temporal correlations in feedback-controlled systems, yet they are also among the most difficult to estimate experimentally. Their evaluation becomes particularly demanding when measurement noise and temporal correlations coexist. By exploiting the finite-memory structure of the control dynamics, we show that these quantities can nevertheless be directly estimated in an experimentally relevant setting. Beyond the specific system studied here, this establishes a practical framework for investigating information encoded in temporal correlations in experimentally accessible regimes.

Importantly, the mechanism underlying the observed crossover is not tied to the specific implementation considered here. Although our analysis is performed within a minimal model in which the measurement outcome is identified with the control variable, the underlying mechanism is generic: measurement noise affects descriptions that rely on detailed trajectory statistics more strongly than those based on instantaneous correlations. We therefore expect the same qualitative behaviour to persist in more complex feedback architectures and measurement schemes. In this sense, our work points to a general limitation of information-theoretic descriptions that rely on increasingly detailed trajectory statistics in realistic experimental settings, where measurement noise progressively limits access to the temporal structure that these descriptions are designed to exploit.

More broadly, our results show that the thermodynamic value of information is not an intrinsic property of a given information measure, but depends on how information is acquired, distributed across trajectories, and ultimately degraded by noise. From this perspective, the hierarchy between information-theoretic descriptions is not universal, but emerges from the physical conditions under which information remains experimentally accessible and thermodynamically relevant.

Finally, our work opens several directions for future research. An important question is whether feedback strategies can be designed to exploit temporal correlations while remaining robust against measurement noise. More generally, understanding how information degradation, temporal correlations, and coarse-graining jointly influence information-theoretic descriptions remains a key challenge for the development of a comprehensive theory of information thermodynamics and for identifying descriptions of information that remain thermodynamically relevant under realistic operating conditions.

\bibliography{Mi-biblioteca-26-may-2026}


\section*{Methods}


\subsection*{Underdamped stochastic dynamics}

Between consecutive measurements, i.e. in the time intervals $(t_n, t_{n+1})$, the particle evolves according to dimensionless underdamped Langevin dynamics, 
\begin{align}
    \dot x &= v, \\
    \dot v &= -\gamma v-(x-cL)+\sqrt{2\gamma}\, \xi(t), 
\end{align}
where $\gamma=1/Q_f$ and $\xi(t)$ denotes unit Gaussian white noise satisfying
\begin{equation}
   \langle \xi(t)\rangle=0, \quad  \langle \xi(t)\xi(t')\rangle=\delta(t-t').
\end{equation}

In the same intervals, the joint probability density $P(\Gamma, c, t)$ evolves according to the Kramers equation
\begin{equation}\label{eq:Kramers-equation-methods}
\partial_t P
=
-\partial_x(vP)
+\partial_v
\left[
\left(
\gamma v+\partial_x V
\right)P
+
\gamma\, \partial_vP
\right].
\end{equation}
For fixed protocol $c$, the stationary equilibrium distribution reads
\begin{equation}
P_{\st}(\Gamma|c)
=
\frac{1}{2\pi}
\exp
\left[
-\frac{1}{2}
\left(
v^2 + (x-cL)^2
\right)
\right].
\end{equation}

Repeated measurements combined with feedback generate a long-time periodic state, 
\begin{equation}
    P(\Gamma, c, t+\Delta t_{\meas})
    =
    P(\Gamma, c, t), 
\end{equation}
which defines the long-time regime analysed throughout this work. The measurement model is detailed in the following.


\subsection*{Measurement model}

The feedback process is described by the conditional probability $\Theta(c|x)$, which gives the probability of applying protocol $c$ when the measured system position is $x$. For perfect measurements, the controller performs a deterministic coarse-graining of phase space, 
\begin{equation}
    \Theta(c|x)
    =
    \mathbf 1_{\chi_c}(x), 
\end{equation}
where $\mathbf 1_{\chi_c}(x)$ denotes the indicator function of the measurement region associated with the outcome $c$. In the binary measurement scheme employed in this work, $\Theta(+1|x)=H(x)$, where $H(\cdot)$ is the Heaviside step function, and $\Theta(-1|x)=1-\Theta(+1|x)=H(-x)$.

Imperfect measurements are modelled through a finite spatial uncertainty $\Delta x$. In that case, the detector only determines whether the particle is located to the right or left of the origin with finite precision. The corresponding conditional probability becomes
\begin{equation}\label{eq:Theta-imperfect-methods}
\Theta(+1|x)=
\begin{cases}
0, & x<-\Delta x, \\
\dfrac{x+\Delta x}{2\Delta x}, & |x|\le\Delta x, \\
1, & x>\Delta x, 
\end{cases}
\end{equation}
and again $\Theta(-1|x)=1-\Theta(+1|x)$.

Perfect measurements therefore correspond to deterministic feedback decisions, whereas imperfect measurements assign finite probabilities to both outcomes for sufficiently small $|x|$. As a consequence of the measurement, the joint probability instantaneously changes in the infinitesimal time interval $(t_n^-, t_n^+)$:
\begin{equation}
    P(\Gamma, c, t_n^+)=P(\Gamma, t_n^-)\Theta(c|x).
\end{equation}
Since measurement does not alter the particle state $\Gamma$, one has $P(\Gamma, t_n^+)=P(\Gamma, t_n^-)\equiv P(\Gamma, t_n)$. The function $\Theta(c|x)$ acquires the interpretation of a post-measurement conditional probability, 
\begin{equation}\label{eq:theta-cn-xn-prob-cond}
    P(c|\Gamma, t_n^+)\equiv \frac{P(\Gamma, c, t_n^+)}{P(\Gamma, t_n^+)}=\Theta(c|x).
\end{equation}

Throughout this work, expressions such as $P(\Gamma_n, c_n)$, $P(\Gamma_n)$, and $P(c_n)$ are used as shorthand notations for $P(\Gamma, c, t_n^+)$, $P(\Gamma, t_n)$, and $P(c, t_n^+)$, respectively. Accordingly, Eq.~\eqref{eq:theta-cn-xn-prob-cond} will often be written as $\Theta(c_n|x_n)$, where $c_n$ denotes the protocol value immediately after measurement at time $t_n$. By contrast, quantities such as $P(\Gamma_n, c_{n-1})$ refer to the pre-measurement state $P(\Gamma, c, t_n^-)$.


\subsection*{Transfer entropy for perfect and imperfect measurements}

In the minimal model considered throughout this work, the measurement outcome coincides with the control variable, $y=c$. Using the transfer-entropy definition introduced in Eq.~\eqref{eq:Ic-Gamma-y-def}, one obtains
\begin{equation}
    \mathcal I_c[\vec\Gamma_k, \vec c_k]
    =
    \left\langle
    \ln
    \frac{
    \Theta(\vec c_k|\vec\Gamma_k)
    }{
    P(\vec c_k)
    }
    \right\rangle, 
\end{equation}
where
\begin{equation}
    \Theta(\vec c_k|\vec\Gamma_k)
    =
    \prod_{n=1}^k
    \Theta(c_n|\Gamma_n).
\end{equation}

For perfect measurements, the conditional entropy vanishes, and the transfer entropy reduces to the entropy of the control sequence introduced in Eq.~\eqref{eq:2nd-law-Ash}, 
\begin{equation}
    \calI_c[\vec\Gamma_k, \vec c_k]
    =
    \calI[\vec c_k].
\end{equation}
For imperfect measurements, however, the conditional entropy remains finite and the transfer entropy becomes
\begin{equation}
    \mathcal I_c[\vec\Gamma_k, \vec c_k]
    =
    \mathcal I[\vec c_k]
    -
    \sum_{n=1}^k
    S(c_n|\Gamma_n), 
\end{equation}
where $S(c_n|\Gamma_n)\equiv -\av{\ln P(c_n|\Gamma_n)}\ge 0$. Measurement uncertainty therefore reduces the transfer entropy relative to the entropy of the control sequence.

The finite conditional entropy generated by measurement noise leads to the qualitatively different behaviour observed in the perfect- and imperfect-measurement regimes. In particular, the ordering relation~\eqref{eq:transfer-unav-ordering}, 
valid for perfect measurements, no longer holds in general in the presence of measurement noise, as illustrated in Fig.~\ref{fig:thermodinamic_deltatm_var}.


\subsection*{Construction of the backward process}

The unavailable information requires the construction of a backward process associated with each forward feedback trajectory. During the forward process, repeated measurements generate a control sequence $\vec c_k=\{c_1, \ldots, c_k\}$. The corresponding backward process is constructed by time reversing this sequence, $\vec c_k^\dg=\{c_k, \ldots, c_1\}$, and subsequently applying it to the system without feedback. The initial distribution of the backward process is the final distribution of the forward dynamics conditioned on the control sequence that defines the backward protocol, $P^R(\Gamma_1^R|\vec{c}_k^{\dg})=P(\Gamma_k=\Gamma_1^R|\vec{c}_k)$.

Measurements are still performed during the backward evolution, generating a backward sequence $\vec c_k^R=\{c_1^R, \ldots, c_k^R\}$, although these measurements no longer affect the externally imposed protocol. The unavailable information is determined from the probability that the backward measurements reproduce the time-reversed forward sequence, namely $\vec c_k^R=\vec c_k^\dg$.

Operationally, the backward experiment was implemented by first recording sufficiently long control sequences in the long-time periodic state and subsequently replaying the corresponding time-reversed control sequences while monitoring the measurement outcomes independently of the applied control. In this way, ensembles of backward trajectories were obtained for each forward trajectory.


\subsection*{Finite-memory estimators}

The entropy and unavailable-information rates are estimated using regularised plug-in estimators~\cite{paninski_estimation_2003, schurmann_entropy_1996}. The estimators are built by expressing both $\calI[\vec c_k]$ and $\calI_u[\vec c_k]$ as sums of per-measurement contributions, namely
\begin{subequations}\label{eq:def-I}
\begin{align}
{\calI[\vec{c}_k]}
&\equiv -\langle \ln P(\vec{c}_k)\rangle
= S(c_1)+\sum_{n=2}^{k} {\calI^{(n)}[\vec{c}_n]}, \\
{\calI^{(n)}[\vec{c}_n]} 
&\equiv  -\av{\ln P(c_n| \vec{c}_{n-1})}
= S(c_n| \vec{c}_{n-1}),
\end{align}
\end{subequations}
and
\begin{subequations}\label{eq:def-Iu}
\begin{align}
\calI_u[\vec{c}_k]
&\equiv
-\av{\ln P^R(\vec{c}_k^{\,R}=\vec{c}_k^{\dg}|\vec{c}_k^{\dg})}
\nonumber\\
&=
-\av{\ln P^R(c_1^R=c_k|\vec{c}_k^{\dg})}
+\sum_{n=2}^k \calI_u^{(n)}[\vec{c}_k],
\\
\calI_u^{(n)}[\vec{c}_k]
&\equiv
-\av{
\ln P^R
\!\left(
c_n^R=c_n^{\dg}
\middle|
\vec{c}_{n-1}^{\,R}=\vec{c}_{n-1}^{\,\dg},
\vec{c}_k^{\,\dg}
\right)
\!}\!.
\end{align}
\end{subequations}

Although the control sequence is non-Markovian, its temporal correlations remain effectively finite-ranged, with a characteristic correlation time of order $\trelax$. As a consequence, the per-measurement contributions defined in Eqs.~\eqref{eq:def-I} and \eqref{eq:def-Iu} become time-translation invariant for sufficiently large $n$,
\begin{equation}\label{eq:convergence-HJ}
\calI^{(n)}[\vec c_n]
=
\mathcal H,
\qquad
\calI_u^{(n)}[\vec c_k]
=
\mathcal J,
\end{equation}
where $\mathcal H$ and $\mathcal J$ are the asymptotic entropy and unavailable-information rates per cycle. Accordingly,
\begin{equation}
\overline{\mathcal I}[\vec c]
=
\lim_{k\to\infty}
\frac{\mathcal I[\vec c_k]}{k}
=
\mathcal H,
\qquad
\overline{\mathcal I_u}[\vec c]
=
\lim_{k\to\infty}
\frac{\mathcal I_u[\vec c_k]}{k}
=
\mathcal J.
\end{equation}

To estimate the asymptotic quantities $\mathcal H$ and $\mathcal J$, we exploit the finite-ranged nature of the temporal correlations. Let us define
\begin{equation}
\vec c_{n-1}^{\,M}
\equiv
\{c_{n-M},\ldots,c_{n-1}\}, \quad n>M,
\end{equation}
which contains the $M$ control actions immediately preceding $c_n$ in the forward process, and analogous chains $\vec{c}_{n-1}^{\,R,M}$ and $\vec{c}_{n-1}^{\dg,M}$ for the backward process. Under a finite-memory approximation, the forward and backward dynamics are described by
\begin{equation}\label{eq:memory-forward-method}
P(c_n|\vec c_{n-1})
\simeq
P(c_n|\vec c_{n-1}^{\,M}),
\end{equation}
and
\begin{equation}\label{eq:memory-back-method}
P^R(c_n^R|\vec c_{n-1}^{\,R},\vec c_n^{\,\dg})
\simeq
P^R(c_n^R|\vec c_{n-1}^{\,R,M},
\vec c_n^{\,\dg,M+1}),
\end{equation}
for large enough $M$. Here,
$\vec c_n^{\,\dg,M+1}
=
\{c_n^{\,\dg},
\vec c_{n-1}^{\,\dg,M}\},
$
so that the same memory depth is retained for both the measurement ($R$) and protocol ($\dg$) sequences. 

The above discussion motivates the introduction of the finite-memory estimators
\begin{subequations}\label{eq:estimator-HJ}
\begin{align}
\mathcal H_M
&\equiv
-\Big\langle
\ln P(c_n|\vec c_{n-1}^{\,M})
\Big\rangle,
\\[1mm]
\mathcal J_M
&\equiv
-\Big\langle
\ln
P^R
\!\left(
c_n^R=c_n^{\,\dg}
\,\middle|\,
\vec c_{n-1}^{\,R,M}
=
\vec c_{n-1}^{\,\dg,M},
\vec c_n^{\,\dg,M+1}
\right)
\Big\rangle,
\end{align}
\end{subequations}
which depend on the retained memory $M$. Once the retained memory exceeds the actual memory of the corresponding process, the estimators are expected to reach plateau values,
\begin{equation}
\mathcal H_M
\simeq
\mathcal H,
\quad
M\ge M_{\cal H}, \qquad
\mathcal J_M
\simeq
\mathcal J,
\quad
M\ge M_{\cal J}.
\end{equation}
The parameters $M_{\cal H}$ and $M_{\cal J}$ denote the effective memory lengths of the forward and backward processes, respectively. Since the memory time is ultimately set by the relaxation dynamics, one expects 
\begin{equation}\label{eq:M0-estimate}
M_{\cal H}(\Delta t_{\meas}),
\;
M_{\cal J}(\Delta t_{\meas})
\propto
\frac{\trelax}{\Delta t_{\meas}}.
\end{equation}
Determining the asymptotic rates $\overline{\calI}[\vec c]$ and $\overline{\calI_u}[\vec c]$ therefore reduces to identifying the plateau regions of $\mathcal H_M$ and $\mathcal J_M$. 

The estimators $\mathcal H_M$ and $\mathcal J_M$ are evaluated as functions of the memory length $M$. As shown in Fig.~\ref{fig:J_H_convergence}, both quantities rapidly approach well-defined plateaux, confirming that temporal correlations in the feedback dynamics are effectively finite-ranged. The location of these plateaux yields direct estimates of the effective memory lengths $M_{\cal H}(\Delta t_{\meas})$ and $M_{\cal J}(\Delta t_{\meas})$. Figure~\ref{fig:J_H_convergence} also supports the scaling relation \eqref{eq:M0-estimate} for $M_{\cal H}$ and $M_{\cal J}$.

Beyond the plateau region, both estimators eventually deviate from their limiting values (crossed-out symbols in Fig.~\ref{fig:J_H_convergence}). This deviation is purely statistical and arises from finite-sample effects: as $M$ increases, the number of possible trajectory blocks grows exponentially as $2^{M+1}$, eventually preventing reliable probability estimation~\cite{ruiz-pino_information_2023}. The final estimates for $\mathcal H$ and $\mathcal J$ are obtained by averaging $\mathcal H_M$ and $\mathcal J_M$ over the plateau region shown in Fig.~\ref{fig:J_H_convergence}. 

Additional details regarding the construction of these estimators, backward-process sampling, regularisation procedures, and statistical analysis are provided in the Supplementary Information.


\subsection*{Evaluation of the Markovian mutual information}

In contrast to the chain-dependent quantities $\mathcal I[\vec c]$ and $\mathcal I_u[\vec c]$, the Markovian mutual-information bound depends only on the instantaneous stochastic process $(\Gamma, c)$ and therefore remains experimentally accessible independently of the memory effects present in the control sequence.

The mutual information between the continuous system state $\Gamma=(x, v)$ and the binary control variable $c\in\{-1, 1\}$ is estimated using the $k$-nearest-neighbour estimator for mixed continuous-discrete variables introduced in Ref.~\cite{ross_mutual_2014}. The corresponding mutual information reads
\begin{equation}
I^{\pm}(\Gamma;c)
=
\sum_{c=\pm1}
P(c)
\int d\Gamma\, P(\Gamma|c, t_n^\pm)
\ln
\frac{P(\Gamma|c, t_n^\pm)}{P(\Gamma, t_n^\pm)}.
\end{equation}

The estimator is applied separately to the pre- and post-measurement states, using the samples $\Gamma_n^-=\Gamma(t_n^-)$ and $\Gamma_n^+=\Gamma(t_n^+)$, respectively. For each sample, the distance to its $k$th nearest neighbour within the corresponding conditional subset is computed, yielding the estimator
\begin{equation}
\hat I^{\pm}(\Gamma;c)
=
\psi(N)+\psi(k)
-
\frac{1}{N}
\sum_{n=1}^N
\left[
\psi(N_{c_n})
+
\psi(m_n)
\right], 
\end{equation}
where $\psi$ denotes the digamma function, $N$ is the total number of samples, $N_{c_n}$ the number of samples belonging to the same conditional subset as $\Gamma_n^\pm$, and $m_n$ the number of points contained within the same neighbourhood in the pooled sample.

Before estimation, both components of $\Gamma$ are normalised using the mean and standard deviation of the pooled sample. Since this corresponds to a global invertible linear transformation applied uniformly to all samples, it leaves the mutual information invariant. Throughout this work, we use $k=3$, corresponding to a standard compromise between local resolution and statistical stability.

The theoretical values are obtained from the stationary solutions of the Kramers dynamics combined with the measurement statistics defined by Eq.~\eqref{eq:Theta-imperfect-methods}. Experimental estimates are computed directly from the measured pre- and post-measurement distributions.

Additional implementation details are provided in the Supplementary Information.


\subsection*{Long-sampling-time limit}

In the limit $\Delta t_{\meas}\gg \trelax$, the system relaxes between consecutive measurements towards the equilibrium distribution associated with the applied protocol. In this regime, both the forward and backward control sequences become effectively Markovian, 
\begin{subequations}
 \begin{align}
P(c_n|\vec c_{n-1})
&=
P(c_n|c_{n-1}), 
\\
P^R(c_n^R|\vec c_{n-1}^R, \vec c_n^{\,\dg})
&=
P^R(c_n^R|c_{n-1}^R, c_n^{\,\dg}).
\end{align}   
\end{subequations}

The entropy and unavailable-information rates then reduce to
\begin{equation}
    \overline{\mathcal I}[\vec c]
    =
    S(c_n|c_{n-1}), 
\end{equation}
and
\begin{equation}
\overline{\mathcal I_u}[\vec c]
=
-
\left\langle
\ln
P^R
(
c_n^R=c_n^{\,\dg}
|
c_{n-1}^R=c_{n-1}^\dg, 
c_n^{\,\dg}
)
\right\rangle, 
\end{equation}
respectively, as shown in the Supplementary Information.

Analytical expressions for the three information-theoretic bounds can therefore be obtained directly from the corresponding forward and backward transition probabilities.

\medskip

Data supporting this study will be available in an open public repository upon acceptance of the manuscript. 

\medskip 

\acknowledgments

We thank C.~Jarzynski for enlightening discussions. AP and NRP acknowledge financial support from Grant PID2024-155268NB-I00 funded by MICIU/AEI/10.13039/501100011033/ FEDER, UE, and also from the applied research and innovation Project PPIT2024-31833, cofunded by EU--Ministerio de Hacienda y Función Pública--Fondos Europeos--Junta de Andalucía--Consejería de Universidad, Investigación e Innovación. NRP acknowledges support from the FPU programme through Grant FPU2021/01764, and also from its mobility programme Grant EST25/00335 that funded the stay at ENS Lyon during which this work was started. LB acknowledges funding by project ANR-22-CE42-0022.

\end{document}